\documentclass[epsfig]{ped}
\usepackage{multicol}
\usepackage{makeidx}
\usepackage{epsfig}
\usepackage{amssymb,amsmath}
\usepackage{citesort}
\usepackage{epic}
\usepackage{eepic}

\newcommand{\vm}{\ensuremath v_{max}}

\begin{document}

\frontmatter          

\pagestyle{headings}  
\addtocmark{random walks} 

\title{Cellular Automaton Approach to Pedestrian Dynamics - Theory}

\author{Andreas Schadschneider}

\institute{Institut f\"ur Theoretische Physik, Universit\"at zu K\"oln\\
  D-50937 K\"oln, Germany\\
  email: as@thp.uni-koeln.de
}

\maketitle              

\begin{abstract}
We present a 2-dimensional cellular automaton model for the simulation
of pedestrian dynamics.  
The model is extremely efficient and allows simulations of large
crowds faster than real time since it includes only nearest-neighbour
interactions. Nevertheless it is
able to reproduce collective effects and self-organization encountered 
in pedestrian dynamics. This is achieved by introducing a so-called
\emph{floor field} which mediates the long-range interactions between 
the pedestrians. This field modifies the transition rates to neighbouring 
cells. It has its own dynamics (diffusion and decay) and can be
changed by the motion of the pedestrians. Therefore the model uses an 
idea similar to chemotaxis, but with pedestrians following a virtual
rather than a chemical trace.

\end{abstract}



\section{Introduction}

Efficient computer simulations of large crowds consisting of hundreds
or thousands of individuals require simple models which nevertheless
provide an accurate description of reality.  One such class of models,
so-called cellular automata (CA), has been studied in
statistical physics for a long time \cite{Wolfram,stauca}.

In CA space, time and state variables are discrete which makes them
ideally suited for high-performance computer simulations.
However, CA modelling differs in several respects from continuum models. 
These are usually based on coupled differential equations which often
can not be treated analytically. One has to solve them numerically and
therefore the equations have to be discretized. 
In general, only space and time variables become 
discrete whereas the state variable is still continuous. One important
point is now that CA are discrete from the beginning and that this
discreteness is already taken into account in the definition of the
model and its dynamics. This allows to obtain the desired behaviour in
a much simpler way. On the other hand, the numerical solution of 
(discretized) differential equations is only accurate in the limit
$\Delta x$, $\Delta t \to 0$. This is different in the CA where
$\Delta x$ and $\Delta t$ are finite and accurate results can be 
obtained since the rules (dynamics) are designed such that
the discreteness is an important part of the model.

In order to achieve complex behaviour in a simple fashion one often
resorts to a stochastic description. A realistic situation seldomly can
be described completely by a deterministic approach. Already  minor
events can lead to a very different behaviour due to the complexity
of the interactions involved. For the problem of pedestrian motion
this becomes evident e.g.\ in the case of a panic where the behaviour
of people seems almost unpredictable. But also for ``normal''
situations a stochastic component in the dynamics can lead to a
more accurate description of complex phenomena since it takes into
account that we usually do not have full knowledge about the state 
of the system and its dynamics. Here one has to keep
in mind that in general some sort of average over different realizations 
of the process (e.g.\ different sequences of random numbers) has
to be taken. Even if there are single realizations which yield
unrealistic behaviour the average process will be a good description
of the real process. Furthermore  a stochastic description allows to
answer questions like ``What is the probability that the evacuation
of this building will take longer than 3 minutes ?'' in a natural way.

In the following we will present a detailed description of the
model and the basic philosophy of our approach. Applications are
presented in Part II \cite{part2}.


\section{Other Modelling Approaches}

During the last decade considerable research has been done on the 
topic of highway traffic using methods from physics
\cite{juelich,tgf97,tgf99,helb,chowd,nagel99,dhrev}.
Cellular automata inspired by the pioneering works
\cite{NagelS,BML} compose by now an important class of models.  
Most studies have been devoted to one-dimensional systems, where 
several analytic approaches exist to calculate or approximate the 
stationary state.

On the other hand, pedestrian dynamics has not been studied as extensively
as vehicular traffic, especially using a cellular automata approach.
One reason is probably its generically two-dimensional nature.
In recent years, continuum models have been most successful in modelling
pedestrian dynamics. An important example are the {\em social force models}
(see e.g.\ \cite{helb,dhrev,social} and references therein). 
Here pedestrians are treated as particles\footnote{In the following
we use ``pedestrian'' and ``particle'' interchangeably.} subject to
long-ranged forces
induced by the social behaviour of the individuals. This leads to
(coupled) equations of motion similar to Newtonian mechanics. There are,
however, important differences since, e.g., in general the third law
(``actio = reactio'') is not fulfilled.

In contrast to the social force models our approach is closer in spirit
to the general strategy of modelling (elementary) forces on a microscopic
level by the exchange of mediating particles which are bosons. It is 
therefore similar to {\em active walker models} \cite{activewalker,trail}
used so far mainly to describe trail formation, chemotaxis 
(see \cite{benjacob} for a review) etc. Here
the walker leaves a trace by modifying the underground on his path.
This modification is real in the sense that it could be measured in
principle. For trail formation, vegetation is destroyed by the walker
and in chemotaxis he leaves a chemical trace. In contrast, in our model
the trace is virtual. Its main purpose is to transform effects of
long-ranged interactions (e.g.\ following people walking some distance
ahead) into a local interaction (with the ``trace''). This allows 
for a much more efficient simulation on a computer.

Cellular automata for pedestrian dynamics have been proposed in
\cite{fukui,nagatani,hubert}.
These models can be considered as generalizations of the 
Biham-Middleton-Levine model for city traffic \cite{BML}.
Most works have focussed on the occurrence of a jamming transition
as the density of pedestrians is increased. All models have only
nearest-neighbour interactions, except for the generalization proposed in
\cite{hubert} which is used for analyzing evacuation processes
on-board passenger ships. The other models use a kind of 
"sublattice-dynamics" which distinguishes between different types
of pedestrians according to their preferred walking direction.
Such an update is not easy to generalize to more complex situations
where the walking direction can change.
To our knowledge so
far most other collective effects encountered empirically 
\cite{helb,dhrev,CrowdFluids,CrowdFluids2,weidmann,panic} have not been
reproduced using these models.
Another discrete model has been proposed earlier by Gipps and 
Marksj\"os \cite{gipps}. This model is somewhat closer in spirit to 
our model than the cellular automata approaches of 
\cite{fukui,nagatani,hubert} since the
transitions are determined by the occupancies of the neighbouring cells.
However, this model can not reproduce all collective effects either.
In \cite{bolay} a discretized version of the social force model
has been introduced. The repulsive potentials by the pedestrians are
stored in a global potential, with pedestrians reacting to the
gradients of this global potential. Although this model is able
to reproduce collective effects it is not flexible enough to treat 
individual reactions to other pedestrians, and collision-avoidance is 
not always guaranteed for velocities greater than 1.


\section{Basic Principles of the Model}
\label{sec_principles}

First we discuss some general principles we took into account in the
development of our model \cite{ourpaper}.
The implementation of the interactions between the pedestrians
uses an idea similar to chemotaxis.
The pedestrians leave a virtual trace which then influences
the motion of other pedestrians. This allows for a very efficient
implementation on a computer since now all interactions are local.
The transition probabilities for all pedestrians just depend
on the occupation numbers and strength of the virtual trace in
his neighbourhood, i.e.\ we have translated the long-ranged spatial
interaction into a local interaction with ``memory''. The
number of interaction terms in other long-ranged models, e.g.\
the social-force model, grows proportionally to the square of
the number of particles whereas in our model it grows only linearly.

The idea of a virtual trace can be generalized to a so-called {\em floor
field}. This floor field includes the virtual trace created by the
pedestrians as well as a static component which does not change with
time. The latter allows to model e.g.\ preferred areas, walls and other 
obstacles. The pedestrians then react to both types of floor fields. 

To keep the model simple, we want to provide the particles with as little 
intelligence as possible and to achieve the formation of complex structures 
and collective effects by means of self-organization.  
In contrast to older approaches we do not make 
detailed assumptions about the human behaviour. Nevertheless
the model is able to reproduce many of the basic phenomena.

The key feature to substitute individual intelligence is the floor
field.  Apart from the occupation number each cell carries an additional
quantity (field) which can be either discrete or continuous.  
This field can have its own dynamics given by 
diffusion and decay coefficients.  

Interactions between pedestrians are repulsive for short distances.
One likes to keep a minimal distance to others in order to avoid
bumping into them. In the simplest version of our model this is taken
into account through hard-core repulsion which prevents multiple
occupation of the cells. For longer distances the interaction is
often attractive. E.g.\ when walking in a crowded area it is usually
advantageous to follow directly behind the predecessor. Large crowds
may also be attractive due to curiosity.

With two particle species moving in opposite directions, each with its
own floor field, effects can be observed which are so far only
achieved by continuous models \cite{social}: lane formation  and
oscillation of the direction of flow at doors.  
We consider this model to be another example for the ability of 
cellular automata to create complex behaviour out of simple rules and 
the great applicability of this approach to all kinds of traffic flow problems.

In contrast to vehicular traffic the time 
needed for acceleration and braking is negligible.
The velocity distribution of pedestrians is sharply peaked
\cite{CrowdFluids}.  These facts naturally lead to a model where
the pedestrians have a maximal velocity $\vm =1$,
i.e.\ only transitions to neighbour cells are allowed.
Furthermore, a greater $\vm$, i.e.\ pedestrians are allowed to move
more than just one cell per timestep, would be harder to implement
in two dimensions, especially when combined with parallel dynamics,
and reduce the computational efficiency.
The number of possible target cells increases quadratically with the
interaction range. Furthermore one has to check whether the path
is blocked by other pedestrians. This might even be ambigious
for diagonal motion and crossing trajectories.
Also higher velocity models lead to timescales which are much too small
(see Sec.\ \ref{sec_def}). 


\section{Definition of the Model and its Dynamics}
\label{sec_def}

The area available for pedestrians is divided into cells of
approximately $40\times 40~cm^2$. This is the typical space occupied
by a pedestrian in a dense crowd \cite{weidmann}.
Each cell can either be empty or occupied by exactly one particle
(pedestrian). 
For special situations it might be desirable to use a finer discretization,
e.g.\ such that each pedestrian occupies four cells instead of one. 

The update is done in parallel for all particles. This introduces
a timescale into the dynamics which can roughly be identified with the
reaction time $t_{\rm reac}$. In the deterministic limit, corresponding
to the maximal possible walking velocity in our model, a single
pedestrian (not interacting with others) moves with a velocity of
one cell per timestep, i.e.\ $40~cm$ per timestep. Empirically the
average velocity of a pedestrian is about $1.3~m/s$ \cite{weidmann}.
This gives an estimate for the real time corresponding to one timestep
in our model of approximately $0.3~sec$ which is of the order of
the reaction time $t_{\rm reac}$, and thus consistent with our
microscopic rules. It also agrees nicely with the time needed to
reach the normal walking speed which is about $0.5~sec$.
This corresponds to at least $v_{max}$ timesteps if the pedestrian
can only accelerate by one unit per timestep. Therefore in
models with large $v_{max}$ a timestep would correspond to a real
time shorter than the smallest relevant timescale. This makes the
model more complicated than necessary and reduces the efficiency
of simulations.


\subsection{Basic Rules}
\label{sec_rules}

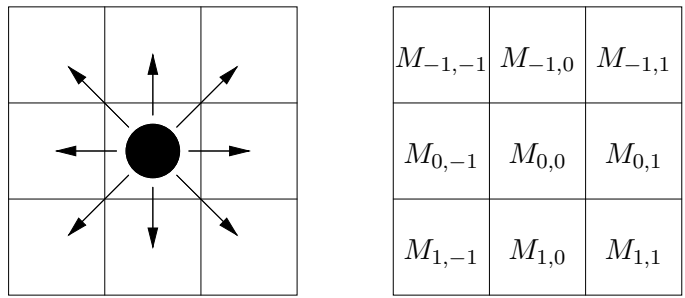
\begin{figure}
  \begin{center}
    \setlength{\unitlength}{0.00083333in}
\begingroup\makeatletter\ifx\SetFigFont\undefined%
\gdef\SetFigFont#1#2#3#4#5{%
  \reset@font\fontsize{#1}{#2pt}%
  \fontfamily{#3}\fontseries{#4}\fontshape{#5}%
  \selectfont}%
\fi\endgroup%
{\renewcommand{\dashlinestretch}{30}
\begin{picture}(4120,1839)(0,-10)
\path(1212,1812)(1212,12)
\path(612,1812)(612,12)
\path(12,612)(1812,612)
\path(12,1812)(1812,1812)(1812,12)
	(12,12)(12,1812)
\path(12,1212)(1812,1212)
\path(3612,1812)(3612,12)
\path(3012,1812)(3012,12)
\path(2412,612)(4212,612)
\path(2412,1812)(4212,1812)(4212,12)
	(2412,12)(2412,1812)
\path(2412,1212)(4212,1212)
\put(912,912){\blacken\ellipse{336}{336}}
\put(912,912){\ellipse{336}{336}}
\path(1137,912)(1512,912)
\path(1137,912)(1512,912)
\blacken\path(1392.000,882.000)(1512.000,912.000)(1392.000,942.000)(1392.000,882.000)
\path(912,687)(912,312)
\path(912,687)(912,312)
\blacken\path(882.000,432.000)(912.000,312.000)(942.000,432.000)(882.000,432.000)
\path(687,912)(312,912)
\path(687,912)(312,912)
\blacken\path(432.000,942.000)(312.000,912.000)(432.000,882.000)(432.000,942.000)
\path(912,1137)(912,1512)
\path(912,1137)(912,1512)
\blacken\path(942.000,1392.000)(912.000,1512.000)(882.000,1392.000)(942.000,1392.000)
\path(1062,1062)(1437,1437)
\path(1062,1062)(1437,1437)
\blacken\path(1373.360,1330.934)(1437.000,1437.000)(1330.934,1373.360)(1373.360,1330.934)
\path(1062,762)(1437,387)
\path(1062,762)(1437,387)
\blacken\path(1330.934,450.640)(1437.000,387.000)(1373.360,493.066)(1330.934,450.640)
\path(762,762)(387,387)
\path(762,762)(387,387)
\blacken\path(450.640,493.066)(387.000,387.000)(493.066,450.640)(450.640,493.066)
\path(762,1062)(387,1437)
\path(762,1062)(387,1437)
\blacken\path(493.066,1373.360)(387.000,1437.000)(450.640,1330.934)(493.066,1373.360)
\put(2712,1437){\makebox(0,0)[b]{\smash{{{\SetFigFont{11}{13.2}{\familydefault}{\mddefault}{\updefault}$M_{-1,-1}$}}}}}
\put(3312,1437){\makebox(0,0)[b]{\smash{{{\SetFigFont{11}{13.2}{\familydefault}{\mddefault}{\updefault}$M_{-1,0}$}}}}}
\put(3912,837){\makebox(0,0)[b]{\smash{{{\SetFigFont{11}{13.2}{\familydefault}{\mddefault}{\updefault}$M_{0,1}$}}}}}
\put(3312,837){\makebox(0,0)[b]{\smash{{{\SetFigFont{11}{13.2}{\familydefault}{\mddefault}{\updefault}$M_{0,0}$}}}}}
\put(2712,837){\makebox(0,0)[b]{\smash{{{\SetFigFont{11}{13.2}{\familydefault}{\mddefault}{\updefault}$M_{0,-1}$}}}}}
\put(2712,237){\makebox(0,0)[b]{\smash{{{\SetFigFont{11}{13.2}{\familydefault}{\mddefault}{\updefault}$M_{1,-1}$}}}}}
\put(3312,237){\makebox(0,0)[b]{\smash{{{\SetFigFont{11}{13.2}{\familydefault}{\mddefault}{\updefault}$M_{1,0}$}}}}}
\put(3912,237){\makebox(0,0)[b]{\smash{{{\SetFigFont{11}{13.2}{\familydefault}{\mddefault}{\updefault}$M_{1,1}$}}}}}
\put(3912,1437){\makebox(0,0)[b]{\smash{{{\SetFigFont{11}{13.2}{\familydefault}{\mddefault}{\updefault}$M_{-1,1}$}}}}}
\end{picture}
}
    \caption{A particle, its possible transitions and the 
associated matrix of preference $M=(M_{ij})$.}
\label{fig_prefs}
  \end{center}
\end{figure}

Each particle is given a preferred walking direction.
From this direction, a $3 \times 3$ \emph{matrix of preferences} is
constructed which contains the probabilities for a move of the
particle. The central element describes the probability for the
particle not to move at all, the remaining 8 correspond to a move to
the neighbouring cells (see Fig.~\ref{fig_prefs}). 
The probabilities can be related to the velocity and the
longitudinal and transversal standard deviations 
(see \cite{ourpaper,diplom} for details).  
So the matrix of preferences contains information about the preferred
walking direction and speed. In principle, it can differ from cell 
to cell depending on the geometry and aim of the pedestrians.
In the simplest case the pedestrian is allowed to move in one direction only 
without fluctuations and in the corresponding matrix of preference only one
element is one and all others are zero.
In the following it is assumed that a matrix of preferences is given 
at every timestep for each pedestrian. These can e.g.\ be
obtained from some model for route selection which assigns certain routes 
to each pedestrian.

This ansatz can easily be extended by fixing the direction of
preference for each cell separately, e.g.\ to handle structures inside
buildings.  Then the particles would use the matrix
belonging to the cell they occupy at a given step.
However, a similar effect can be obtained much simpler by introducing
a second floor field (see Sec.~\ref{sec_floor}).

\begin{figure}
  \begin{center}
    \setlength{\unitlength}{0.00083333in}
\begingroup\makeatletter\ifx\SetFigFont\undefined%
\gdef\SetFigFont#1#2#3#4#5{%
  \reset@font\fontsize{#1}{#2pt}%
  \fontfamily{#3}\fontseries{#4}\fontshape{#5}%
  \selectfont}%
\fi\endgroup%
{\renewcommand{\dashlinestretch}{30}
\begin{picture}(8173,2439)(0,-10)
\put(1212,1812){\blacken\ellipse{300}{300}}
\put(1212,1812){\ellipse{300}{300}}
\put(612,612){\blacken\ellipse{300}{300}}
\put(612,612){\ellipse{300}{300}}
\put(3912,1212){\blacken\ellipse{300}{300}}
\put(3912,1212){\ellipse{300}{300}}
\put(3312,612){\blacken\ellipse{300}{300}}
\put(3312,612){\ellipse{300}{300}}
\put(6312,1812){\blacken\ellipse{300}{300}}
\put(6312,1812){\ellipse{300}{300}}
\put(6312,1212){\blacken\ellipse{300}{300}}
\put(6312,1212){\ellipse{300}{300}}
\path(1212,1587)(1212,1212)
\path(1212,1587)(1212,1212)
\blacken\path(1182.000,1332.000)(1212.000,1212.000)(1242.000,1332.000)(1182.000,1332.000)
\path(762,762)(1137,1137)
\path(762,762)(1137,1137)
\blacken\path(1073.360,1030.934)(1137.000,1137.000)(1030.934,1073.360)(1073.360,1030.934)
\path(312,2412)(312,12)
\path(912,2412)(912,12)
\path(1512,2412)(1512,12)
\path(12,2112)(1812,2112)
\path(1812,1512)(12,1512)
\path(12,912)(1812,912)
\path(1812,312)(12,312)
\path(2712,2112)(4512,2112)
\path(4512,1512)(2712,1512)
\path(2712,912)(4512,912)
\path(4512,312)(2712,312)
\path(3612,2412)(3612,2112)
\path(3612,2412)(3612,2112)
\path(3612,1512)(3612,12)
\path(3612,1512)(3612,12)
\path(3012,2412)(3012,2112)
\path(3012,2412)(3012,2112)
\path(4212,2412)(4212,2112)
\path(4212,2412)(4212,2112)
\path(4212,1512)(4212,12)
\path(4212,1512)(4212,12)
\path(3012,1512)(3012,12)
\path(3012,1512)(3012,12)
\path(5112,2112)(6912,2112)
\path(6912,1512)(5112,1512)
\path(5112,912)(6912,912)
\path(6912,312)(5112,312)
\path(6012,2412)(6012,912)
\path(6012,2412)(6012,912)
\path(6012,312)(6012,12)
\path(6012,312)(6012,12)
\path(6612,12)(6612,312)
\path(6612,12)(6612,312)
\path(5412,12)(5412,312)
\path(5412,12)(5412,312)
\path(5412,912)(5412,2412)
\path(5412,912)(5412,2412)
\path(6612,2412)(6612,912)
\path(6612,2412)(6612,912)
\put(1212,537){\makebox(0,0)[b]{\smash{{{\SetFigFont{10}{12.0}{\familydefault}{\mddefault}{\updefault}$M_{-1,1}^{(2)}$}}}}}
\put(3612,1737){\makebox(0,0)[b]{\smash{{{\SetFigFont{10}{12.0}{\familydefault}{\mddefault}{\updefault}$p_1 = \frac{M_{1,0}^{(1)}}{M_{1,0}^{(1)}+M_{-1,1}^{(2)}}$}}}}}
\put(6012,537){\makebox(0,0)[b]{\smash{{{\SetFigFont{10}{12.0}{\familydefault}{\mddefault}{\updefault}$p_2 = \frac{M_{-1,1}^{(2)}}{M_{1,0}^{(1)}+M_{-1,1}^{(2)}}$}}}}}
\put(2262,1137){\makebox(0,0)[b]{\smash{{{\SetFigFont{10}{12.0}{\familydefault}{\mddefault}{\updefault}becomes}}}}}
\put(4812,1137){\makebox(0,0)[b]{\smash{{{\SetFigFont{10}{12.0}{\familydefault}{\mddefault}{\updefault}or}}}}}
\put(612,1737){\makebox(0,0)[b]{\smash{{{\SetFigFont{10}{12.0}{\familydefault}{\mddefault}{\updefault}$M_{1,0}^{(1)}$}}}}}
\end{picture}
}
    \caption{Solving conflicts according to the relative probabilities for
the case of two particles with matrices of preference $M^{(1)}$ and
$M^{(2)}$.}
\label{fig_conflict}
  \end{center}
\end{figure}
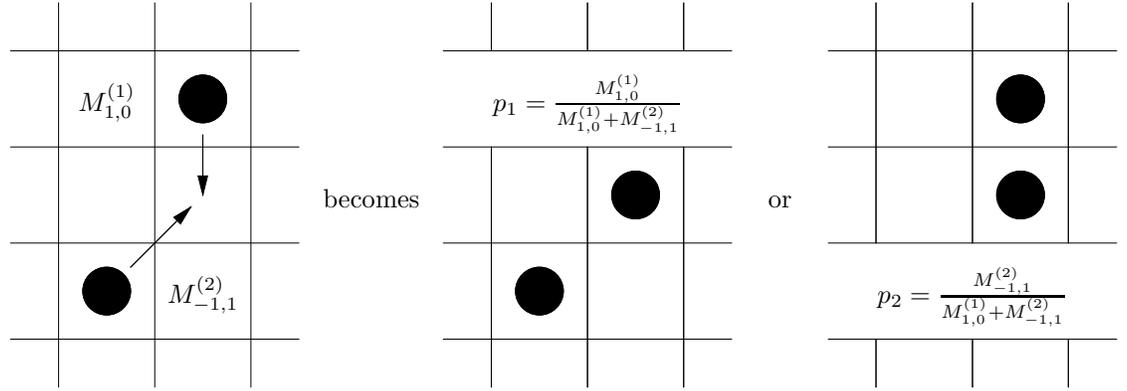

In each update step for each particle a desired move is chosen
according to these probabilities. This is done in parallel for
all particles. If the target cell is occupied, the
particle does not move.  If it is not occupied, and no other particle
targets the same cell, the move is executed.  If more than one
particle share the same target cell, one is chosen according to the
relative probabilities with which each particle chose their target.
This particle moves while its rivals for the same target keep their
position (see Fig.~\ref{fig_conflict}).

The rules presented up to here are a straightforward generalization of the
CA rules used so far for the description of traffic flow
\cite{fukui,nagatani}.
The main difference is that in principle transitions in all directions
are possible and each pedestrian $j$ might have her own preferred
direction of motion characterized by a matrix of preferences $M^{(j)}$.
The only interaction between particles taken into account so far is 
hard-core exclusion.


\subsection{Floor Field}
\label{sec_floor}

In order to reproduce certain collective phenomena it is necessary to
introduce further longer-ranged interactions. In some continuous models
this is done using the idea of a social force \cite{helb,dhrev,social}.
Here we present a different approach. 
Since we want to keep the model as simple as possible we try to avoid
using a long-range interaction explicitly. Instead we introduce the concept
of a {\em floor field} which is modified by the pedestrians and which in turn
modifies the transition probabilities. This allows to take into
account interactions between pedestrians and the geometry of the system
(building) in a unified and simple way without loosing the advantages 
of local transition rules. The floor field modifies the transition
probabilities in such a way that a motion into the direction of larger
fields is preferred. 

The floor field can be thought of as a second grid of cells underlying
the grid of cells occupied by the pedestrians. It can be discrete or
continuous.  
As already explained in Sec.\ \ref{sec_principles} we distinguish 
two types of fields which will be called static and dynamic floor fields,
respectively.

The {\em dynamic floor field} $D$  is just the virtual trace left by the
pedestrians (see Sec.\ \ref{sec_principles}). It is modified by the presence
of pedestrians and has its own dynamics, i.e.\ diffusion and decay.
Usually the dynamic floor field is used to model a (``long-ranged'')
attractive interaction
between the particles. Each pedestrian leaves a ``trace'', i.e.\ the
floor field of occupied cells is increased. 
Explicit examples where such an interaction is relevant are given in 
Part II \cite{part2}. The dynamic floor field is also subject to diffusion
and decay which leads to a dilution and finally the vanishing of the
trace after some time.

The {\em static floor field} $S$ does not evolve with time and is not
changed by the presence of pedestrians. Such a field can be used to
specify regions of space which are more attractive, e.g.\ an emergency
exit (see the example in \cite{part2}) or shop windows. This
has an effect similar to a position-dependent matrix of preference, but
is much easier to realize.

The {\em transition probability} $p_{ij}$ in direction $(i,j)$
(see Fig.~\ref{fig_prefs}) now depends on four contributions:
\begin{itemize}
\item[(i)] the matrix of preference $M_{ij}$ which contains the information
about the aim and average velocity of the pedestrian.
\item[(ii)] the value $D_{ij}$ of the dynamic floor field at the
target cell. This contribution takes into account the effects of the motion
of the other pedestrians. In many applications (see \cite{part2}) it is
attractive to ``follow the crowd'', i.e.\ transitions in directions $(i,j)$
with a large value of the dynamic floor field are preferred.
\item[(iii)] the value $S_{ij}$ of the static floor field. It allows to
model effects of the geometry. E.g.\ in a corridor it is usually less
attractive to walk close to the walls. Such an effect can be incorporated
in a static floor field which decreases near the walls.
\item[(iv)] the occupation number $n_{ij}$ of the target cell. A motion in
direction $(i,j)$ is only allowed if the target cell is empty ($n_{ij}=0$)
and forbidden if it is already occupied ($n_{ij}=1$).
\end{itemize}

One simple possibility to take into account all contributions (i)--(iv) 
is to define the transition probability in direction $(i,j)$ by
\begin{equation}
p_{ij}=NM_{ij}D_{ij}S_{ij}(1-n_{ij}).
\label{transprob}
\end{equation}
$N$ is a normalization factor to ensure $\sum_{(i,j)}p_{ij}=1$
where the sum is over the nine possible target cells.
There are also slightly more general forms of the transition probabilities 
which have been studied in \cite{ourpaper,diplom}.

Since the total transition probability is proportional to the dynamic
floor field it becomes more attractive to follow in the footsteps of
other pedestrians. This effect competes with the preferred walking direction
specified by $M_{ij}$ and the effects of the geometry encoded in $S_{ij}$.
The relative influence of the contributions (i)--(iii) is controlled by
coupling parameters. These depend on the situation to be studied.
Consider for example a situation where people want to leave a large room
(see \cite{part2}). Normal circumstances, where everybody is able
to see the exit, can be modelled by solely using a static floor field
which decreases radially with the distance from the door. Since transitions
in the direction of larger fields are more likely this will automatically
guarantee that everybody is walking in the direction of the door.
If, however, the exit can not be seen by everybody, e.g.\ in a smoke-filled
room or in the case of failing lights, people will try to follow others
hoping that they know the location of the exit. In this case the coupling
to the dynamic floor field is much stronger and the static field has a
considerable influence only in the vicinity of the door. This example
will be studied in more detail in \cite{part2}.


\subsection{Dynamics of the Floor Field}
\label{dynfloor}

In contrast to the static floor field $S$ the dynamic floor field $D$
is changed by the motion of pedestrians. Furthermore it is subject to
diffusion and decay. Its dynamics consists of three steps:
\begin{itemize}
\item[(a)] If a pedestrian leaves a cell $(x,y)$ the dynamic floor field
$D_{xy}$ corresponding to this cell is increased by $\Delta D_{xy}$. 
The increment $\Delta D_{xy}$ is a parameter of the model and can either 
be discrete or continuous.
\item[(b)] To model the diffusion, a certain amount of the
field is distributed among the neighbouring cells.
\item[(c)] To model the decay of the field, the field strength is reduced
by a decay constant $\delta$.
\end{itemize}
In (a) the virtual trace left by the motion of the pedestrians is
created. (b) is necessary because pedestrians do not necessarily follow
exactly in the footsteps of others. Diffusion leads to broadening and
dilution of the trace. (c) implies that the lifetime of the trace is
finite and that it will vanish after some time. Diffusion and decay of
the dynamic field lead to an effective interaction strength between
the pedestrians which decays exponentially with the distance \cite{diplom}.

In \cite{ourpaper} we have introduced two variants of the floor field,
a discrete and a continuous one. In the discrete case the field strength
$D_{xy}$ can be interpreted as the number of bosonic particles (``bosons'')
at the cell $(x,y)$. In (a) the number of bosons is increased by one.
In (b) bosons can move with probability $\gamma$ to neighbouring cells
and in (c) bosons are removed with probability $\alpha$. In the
continuous case the dynamics in (b) and (c) is described by a 
diffusion-decay equation
\begin{equation}
  \frac{\partial D}{\partial t} = d \cdot \Delta D - \delta \cdot D
\label{eq_diffu}
\end{equation}
where $d$ is the diffusion constant and $\delta$ the decay constant.
Details can be found in \cite{ourpaper,diplom}.


\subsection{Summary of the Update Rules}

The update rules of the full model including the interaction with the
floor fields then have the following structure:
\begin{enumerate}
\item The dynamic floor field $D$ is modified according to its diffusion 
and decay rules (see Sec.~\ref{dynfloor}).
\item For each pedestrian, the transition probabilities $p_{ij}$
for a move to an unoccupied neighbour cell $(i,j)$ are determined by 
the matrix of preferences and the local dynamic and static floor fields, 
e.g. $p_{ij}\propto M_{ij}D_{ij}S_{ij}$ (see Sec.~\ref{sec_floor})..
\item Each pedestrian chooses a target cell based on the probabilities
of the transition matrix $P=(p_{ij})$.
\item The conflicts arising by any two or more pedestrians attempting to
  move to the same target cell are resolved, e.g.\ using the procedure
described in Sec.~\ref{sec_rules}.
\item The pedestrians which are allowed to move execute their step.
\item The pedestrians alter the dynamic floor field $D_{xy}$ of the cell
$(x,y)$ they occupied before the move (see Sec.~\ref{dynfloor}).
\end{enumerate}
These rules have to be applied to all pedestrians at the same time
(parallel dynamics).
This introduces a timescale into the dynamics which  corresponds
to approximately $0.3~sec$ of real time. This allows e.g.\ to translate
evacuation times measured in computer simulations into real times.

\section{Conclusions}

We have introduced a stochastic cellular automaton to simulate
pedestrian behaviour.  We focused here on the general concept.
The effects which can be observed with the basic approach will be 
presented in \cite{part2} together with simple applications.

The key mechanism is the introduction of a floor field
which acts as a substitute for pedestrian intelligence and leads to
collective phenomena. This floor field makes it possible to 
translate spatial long-ranged interactions into non-local interactions
in time. The latter can be implemented much more efficiently on a 
computer. Another advantage is an easier treatment of complex
geometries. In models with long-range interactions, e.g.\ the social-force
models, one always has to check explicitly whether pedestrians are 
separated by walls in which case there should be no interaction between
them.

The general idea in our model is similar to chemotaxis. However,
the pedestrians leave a virtual trace rather than a chemical one.
This virtual trace has its own dynamics (diffusion and decay) which
e.g.\ restricts the interaction range (in time). It is
realized through a dynamical floor field which allows to give the
pedestrians only minimal intelligence and to use local interactions.
Together with the static floor field it offers the possibility
to take different effects into account in a unified way, e.g.\
the social forces between the pedestrians or the geometry of the
building.

In Part II \cite{part2} we will demonstrate that the approach indeed is
able to reproduce the known collective effects and self-organization
phenomena. Therefore the model is a good starting point for realistic 
applications \cite{ourpaper,part2}.

The model can also be applied to more complex geometries
and various characteristics of a crowd can be simulated without major
changes. So it should be possible to study the effects of panic
(see \cite{panic} and references therein). In \cite{part2} we show
results for simple evacuation simulations.

The description of pedestrians using a cellular automaton approach
allows for very high simulation speeds. Therefore, we have the
possibility to extract the complete statistical properties of our
model using Monte Carlo simulations. 

Finally it should be emphasized that we have presented only the basic
ideas of the approach. For realistic applications modifications might
be appropriate, e.g.\ smaller cell sizes etc. One can also introduce
more than just one species of pedestrians (e.g.\ two groups moving in
opposite directions). In this case each species interacts with its own 
floor field.
In the simplest case these fields are independent from each other.




\end{document}